\documentclass[prl,twocolumn
]{revtex4-1}

\usepackage{lipsum}  
\usepackage{amsmath}
\usepackage{amsfonts}
\usepackage{amssymb}
\usepackage{comment}
\usepackage{graphicx}% Include figure files
\usepackage{dcolumn}% Align table columns on decimal point
\usepackage{bm}% bold math
\usepackage{float}
\usepackage{appendix}
\usepackage{tcolorbox}

\begin{document}

\preprint{APS/123-QED}

\title{Cost of diffusion: nonlinearity and giant fluctuations}
%Diffusion enhances cost fluctuations
%The cost of diffusion: nonlinearity and giant fluctuations
%Nonlinearity-induced cost fluctuations in diffusion

%Nonlinearity-induced cost fluctuations for random walks

\author{Satya N. Majumdar$^1$, Francesco Mori$^2$, Pierpaolo Vivo$^3$}
 \affiliation{$^1$ LPTMS, CNRS, Univ. Paris-Sud, Université Paris-Saclay, 91405 Orsay, France\\ $^2$  Rudolf Peierls Centre for Theoretical Physics,
University of Oxford, Oxford, United Kingdom\\ $^3$ Department of Mathematics, King’s College London, London WC2R 2LS, United Kingdom}%Lines break automatically or can be forced with \\

\date{\today}

\begin{abstract}
% \lipsum[1]
We introduce a simple model of diffusive jump process where a fee is charged for each jump. The nonlinear cost function is such that slow jumps incur a flat fee, while for fast jumps the cost is proportional to the velocity of the jump. The model -- inspired by the way taxi meters work -- exhibits a very rich behavior. The cost for trajectories of equal length and equal duration exhibits giant fluctuations at a critical value of the scaled distance travelled. Furthermore, the full distribution of the cost until the target is reached exhibits an interesting ``freezing'' transition in the large-deviation regime. All the analytical results are corroborated by numerical simulations. Our results also apply to elastic systems near the depinning transition, when driven by a random force.
\end{abstract}

\maketitle

%\tableofcontents

%\lipsum[2-4]
\noindent\textit{Introduction - }For more than a century, simple stochastic processes like the random walk have been successfully used to model a variety of phenomena across disciplines \cite{berg_book,biology,Bouchaud05}. For instance, the motion of bacteria in space \cite{berg_book,biology} and the evolution of the price of a stock in finance \cite{Bouchaud05} can be approximated as a sequence of jumps between \emph{states}, which take place according to some probabilistic rule. 

In many different contexts, it is natural to associate a (possibly non-linear) cost -- or a reward -- to the ``change of state'' of a stochastic process -- often with unexpected or paradoxical consequences. For instance, the energy consumption
of bacteria changes depending on the environment they move in \cite{berg_book}. Wireless devices absorb different amounts of energy when they switch between activity states (`off', `idle', `transmit' or `receive') \cite{wireless,wireless2}. In biochemical reactions, it is often convenient to label secondary reaction products as cost/reward of an underlying primary process \cite{stochasticreaction} for bookkeeping purposes. The \emph{bonus-malus} vehicle insurance premium changes depending on the number of claims made in the previous year \cite{markovreward4}. In software development, the so-called ``technical debt'' is the cost of additional rework caused by prioritizing an easy solution now instead of a better design approach that would delay the release of the product \cite{technicaldebt}. In a variety of situations where random factors are present that affect the change of state of a system, computing the total cost (or reward) of a trajectory may prove very challenging. 

In Mathematics and Engineering, stochastic processes with associated costs have been investigated in the framework of Markov reward models \cite{howard_book,markovreward,rew}. Recently, the joint distribution of displacement and cost has also been investigated in Ref.~\cite{costlevy} for random walks in a random environment until a first-passage event. Moreover, optimal control theory has been applied in Ref.~\cite{debruyne21} to minimize the cost of random walks with resetting. However, the impact of a nonlinear cost function on the cost fluctuations, both in the typical and in the large deviation regime, remains largely unexplored.

An everyday example where nonlinear costs lead to unexpected consequences is that of taxi fares. Indeed, taxi rides in a busy city typically consist of a mixture of fast excursions, and slow steps due, e.g., to congestion or traffic lights. The fare charged to a passenger is automatically computed by the taxi meter, which follows a fairly universal and simple recipe \cite{Eastaway}. Each city council determines a \emph{changeover speed} $\eta_c$ -- based on a statistical analysis of the typical local traffic conditions. If the taxi moves faster than $\eta_c$, the meter ticks according to the \emph{space} covered, while if the taxi moves slower than $\eta_c$, the meter ticks according to the \emph{time} elapsed. This way, the driver gets compensated even when the taxi barely moves due to heavy traffic. For example, according to London's Tariff I rate \cite{TFL} the meter should charge $20$ pence for every 105.4 metres covered, or 22.7 seconds elapsed (whichever is reached first). One of the surprising consequences of the non-linear nature of the taxi fare structure is the so-called \emph{taxi paradox} \cite{Eastaway}, whereby two taxis starting together from $A$ and arriving together at $B$ may charge very different fares depending on their individual patterns of slow vs. fast chunks in their trajectories.

In this Letter, we study a simple but general model of diffusion, inspired by the taxi paradox, where the nonlinear nature of the costs associated to each jump gives rise to a rich and nontrivial behavior. In particular, we will consider two scenarios: fixing both the total distance $X$ and the number $n$ of steps (Ensemble (i)), or fixing the target location $L$ but allowing the number of steps to get there to fluctuate (Ensemble (ii)). In Ensemble (i), the cost has a finite and non-monotonic variance, which is maximal at some critical value of the scaled distance $X/n$. In Ensemble (ii), we show that the cost variance displays a rich behavior when changing the speed threshold $\eta_c$, including `giant' fluctuations of the total cost. In this latter setting, the large deviations of the hitting cost displays an unexpected `freezing' transition in the low-cost regime. Our results show that associating a nonlinear cost to the evolution of a random walk leads to very rich and unexpected phenomena.

\begin{figure}[t]
    \centering
    \includegraphics[scale = 0.6]{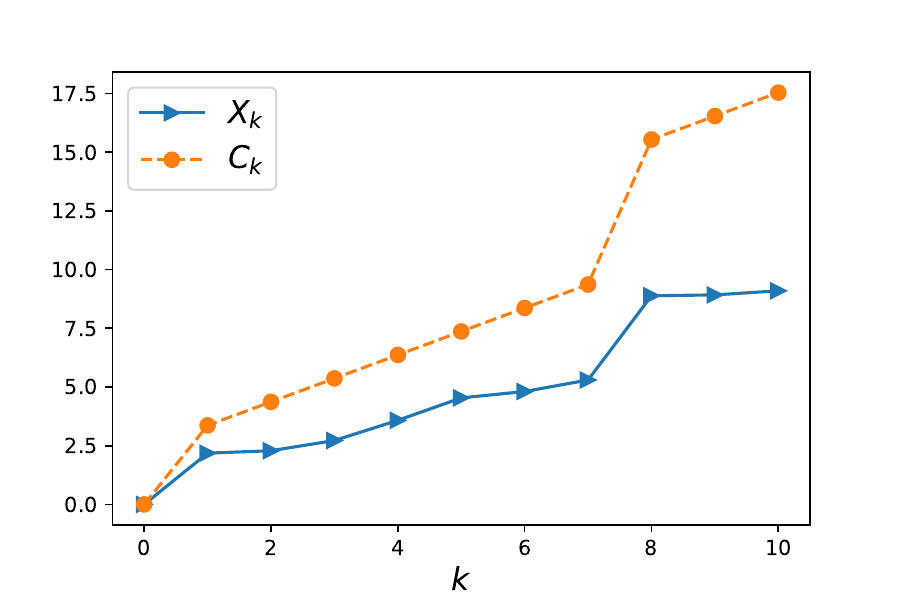}
    \caption{Typical realization of a random walk $X_k$ with cost function $h(\eta)=1+2(\eta-1)\theta(\eta-1)$. The cost $C_k$ up to step $k$ is a nonlinear function of the random jumps.}
    \label{fig:jump}
\end{figure}

\textit{The model - }
Consider a one-dimensional walker whose position $X_n$ at discrete time $n$ evolves according to
\begin{equation}
    X_n = X_{n-1}+\eta_n
\end{equation}
starting from the origin $X_0=0$, with $\eta_n$ drawn independently from a probability density function (pdf) $p(\eta)$ with positive support. To each jump, we associate a cost $C_n$ that increases according to the law
\begin{equation}
    C_n = C_{n-1}+h(\eta_n)
\end{equation}
where $h(\eta)$ is a function of $\eta$. The final position reached after $n$ steps is $X=\sum_{k=1}^n\eta_k$, and the total cost due is $C=\sum_{k=1}^n h(\eta_k)$. For a typical realization of the process, see Fig.~\ref{fig:jump}. Clearly, $X$ and $C$ are correlated random variables, whose joint statistics is of interest here. 

We further assume that the jumps are positive and exponentially distributed with mean value $\mu$, i.e., that $p(\eta)=\exp(-\eta/\mu)/\mu$. For simplicity, we set $\mu=1$. Inspired by the taxi paradox described above, we consider the nonlinear cost function $h(\eta)=1+b(\eta-\eta_c)\theta(\eta-\eta_c)$, with $b>0$ a positive constant, and $\theta(x)$ the Heaviside step function. This function $h(\eta)$ is such that jumps shorter than the critical size $\eta_c$ in one unit of time (\emph{slower} jumps) incur a unit fee, whereas longer (\emph{faster}) jumps are more costly, with the fee being proportional to the length (velocity) of the jump. Thus, in our model there are two parameters, $b$ and $\eta_c$.

\begin{figure}[t]
    \centering
 \includegraphics[scale = 0.56]{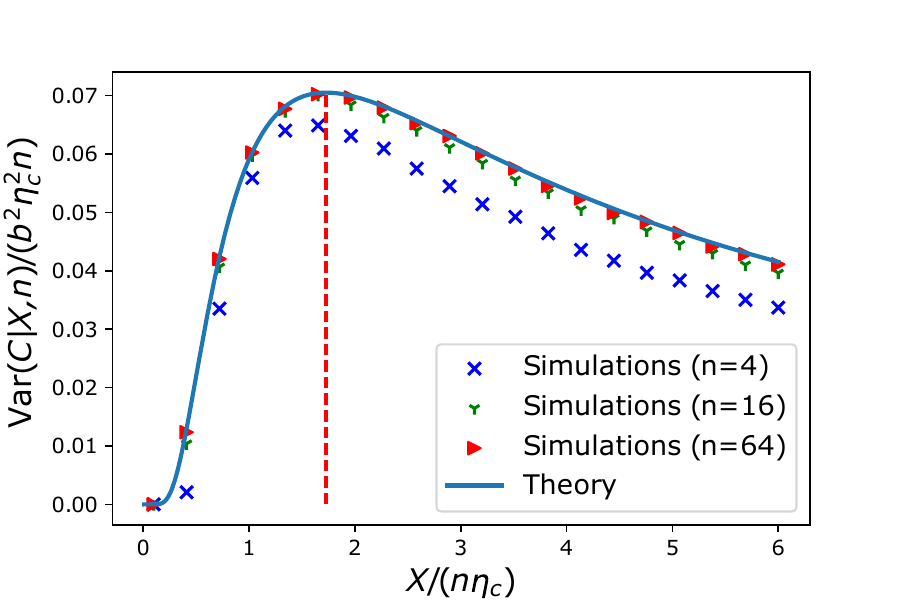}
    \caption{Scaled cost variance $\operatorname{Var}(C|X,n)/(b^2\eta_c^2 n)$ conditioned on the final position $X$ as a function of $X/(n\eta_c)$. The continuous blue line corresponds to the analytical scaling function in Eq. \eqref{eqscalingF} (valid for $n\to \infty$). The symbols display the results of numerical simulations with $10^5$ samples and different values of $n$. The vertical dashed line highlights the maximum at $X/(n\eta_c)\approx1.72724\ldots$.}
    \label{fig:Fcurve}
\end{figure}

\textit{Main results (Ensemble (i)) - } Due to the nonlinear nature of the cost function, even after fixing the total distance $X$ and the number of steps $n$, the total cost $C$ remains random, as expressed by the taxi paradox. For large $n$, the average cost grows with the total distance $X$ as 
\begin{equation}
\langle C\rangle_{X,n}\approx n+b\eta_c nH\left(\frac{X}{n\eta_c}\right)\,,
\label{eq:avg_scaling}
\end{equation}
where $H(y)=ye^{-1/y}$. To quantify the cost fluctuations around the average, we first compute the cost variance $\operatorname{Var}(C|X,n)$ conditioned on the value of $X$ after exactly $n$ steps. In particular, in the late-time limit $n\to\infty$, $X\to\infty$ with $y=X/(n\eta_c)$ fixed, we find that the variance takes the scaling form
\begin{equation}
    \operatorname{Var}(C|X,n)\approx b^2 \eta_c^2 n F\left(\frac{X}{n\eta_c}\right)\,,
    \label{eq:var_scaling}
\end{equation}
where
\begin{equation}
F(y)= e^{-2/y} \left(2 e^{1/y} y^2-2 y^2-2 y-1\right)\label{eqscalingF}
\end{equation}  
is a positive, non-monotonic scaling function. This scaling function is shown in Fig.~\ref{fig:Fcurve} and it is in perfect agreement with numerical simulations. Note that the variance has a single maximum at the point $y^\star\approx 1.72724\ldots$. This scaling function has asymptotic behaviors $F(y)\approx2 y^2 e^{-1/y}$ for small $y$ and $F(y)\approx \frac{1}{3y}$ for large $y$.

To understand the non-monotonic behavior of the variance, consider two opposite limits: $X\gg n\eta_c$ (that is $y$ large) and $X\ll n \eta_c$ ($y$ small). In the former case, when $X$ is fixed to be large,
in a typical trajectory, most of the jumps are big, i.e, the configuration is dominated
by ``space-like'' runs only (i.e., $\eta_i>\eta_c$) and hence the cost can be written as $C=\sum_{i}(1+b(\eta_i-\eta_c))\approx n+b(X-n\eta_c)$. Consequently, since both $n$ and $X$ are fixed the fluctuations of $C$ are severely constrained for large $X$. This decrease in the cost fluctuations is hence consequence of the linearity of $h(\eta)$ for large $\eta$ (see \cite{supplemental} for a discussion on the nonlinear case). In the latter case $X\ll n \eta_c$, a typical trajectory is dominated by only ``time-like'' runs and once again the cost fluctuations from one trajectory to another are expected to be small, as the precise length of each jump will not significantly alter the fee charged per jump. Thus these two `phases' are like `pure' phases. As one increases the `control parameter' $X/(n\eta_c)$, one first crosses over
from a `time-like' pure phase
to a `mixed' phase, characterized by a larger entropy (i.e. a larger number of possible arrangements of individual jumps eventually landing to the same final spot $X$ after $n$ jumps). Upon further increasing $X/(n\eta_c)$, the conditional variance undergoes a second crossover
from the `mixed' phase to a `space-like' pure phase. At the special value $X/(n\eta_c)\approx 1.72724...$, the cost fluctuations are maximal (the most unfair scenario for taxi passengers).

\textit{Physical applications - } Nonlinear functions similar to $h(\eta)$, 
composed of a constant and a linear part, naturally emerge in disparate areas of 
physics. An elementary example is that of static friction: in order to move a block 
in contact with a substrate one has to overcome a threshold force $\eta_c$ due to 
static adhesion. Then, applying a force $\eta$ for a fixed time interval $\Delta t$, 
the velocity of the block is given by $h(\eta)=b\,(\eta-\eta_c)\theta(\eta-\eta_c)$ 
where $b$ now depends on the block mass and $\Delta t$~\cite{VMT2012}. The function 
$h(\eta)$ is the velocity-force characteristic describing the response of the system 
to an applied force and, up to a global shift by $1$, is identical to the cost 
function per unit time in our taxi model. A natural question is: what is the average 
response when the block is subject to a {\em random applied force} $\eta$ drawn from, 
say, $p(\eta)=e^{-\eta}$? To measure this average (over random force) response, one 
needs to repeat the experiment $n$ times, by applying a random force $\eta_i$ drawn 
independently for each sample $i$ from $p(\eta)$. Then $X_n/n= (1/n)\,\sum_{i=1}^n 
\eta_i$ is precisely the mean force per sample, and $C_n/n= (1/n)\sum_{i=1}^n 
h(\eta_i)$ is the mean velocity of the block per sample. Thus, the number of steps 
$n$ in the taxi problem plays the role of the number of samples here. Consequently, 
for large $n$, the scaling function $H(y)= y\, e^{-1/y}$ in Eq. 
(\ref{eq:avg_scaling}) describes precisely the average response characteristic, while 
$F(y)$ in Eqs. \eqref{eq:var_scaling} and \eqref{eqscalingF} describe the 
fluctuations of the response around its average.

More generally, our results can be extended to a wide variety of disordered 
systems when an extended object/manifold such as an elastic string or a polymer is 
driven by a {\em random force} $\eta$ in a spatially inhomogeneous medium. These 
systems undergo a depinning transition when a force $\eta$ is applied: below the 
depinning threshold $\eta_c$, the manifold is pinned by the disorder and its velocity 
vanishes, while above the threshold, the velocity-force relation follows a power-law 
scaling $h(\eta)\propto (\eta-\eta_c)^{\beta}$ with the depinning exponent $\beta > 
0$~\cite{pascal2000,Duemmer2005,Reichhardt2016}. For example, when a DNA chain 
translocates through a nanopore by applying a pulling force via optical tweezer, the 
exponent $\beta\approx 1$~\cite{menais18}, while for a harmonic elastic string in 
$(1+1)$-dimensions one gets $\beta\approx 0.33$~\cite{Duemmer2005}. Other examples 
include vortices in type-II superconductors~\cite{Blatter1994} and colloidal crystals 
\cite{Pertsinidis2008}. To analyse the velocity-force characteristic for such an 
elastic string driven by a {\it random force}, we need to generalise our method 
presented above for $\beta=1$ to $h(\eta)= b\,(\eta-\eta_c)^{\beta}$ for arbitrary 
$\beta>0$. In \cite{supplemental}, we have computed exactly both the average 
velocity-force response characteristic and its fluctuations for arbitrary $\beta$. 
Our results show that the associated scaling functions $H_{\beta}(y)$ and 
$F_\beta(y)$ depend continuously on $\beta$.

\textit{Main results (Ensemble (ii)) - }It is also natural to estimate the distribution $P(C|L)$ of the \emph{hitting cost} to be paid to reach a given location $L$, irrespective of the time required. First-passage or hitting properties  \cite{redner_book,Brayreview,BrownianFunctionals} are important in several applications, from chemical reactions \cite{hanggi90} to insurance policies \cite{avram08}. Note that in this second setting, the number $n$ of steps is a random variable. We find that for large $L$ the distribution of $C$ takes the large-deviation form
\begin{equation}
P(C|L)\sim \exp(-L\Phi(C/L))\,,
\label{eq:large_deviation}
\end{equation}
where the rate function reads
\begin{equation}
    \Phi(z)=\max_s\left[-sz+1+u(s)\right]\label{legendrephi}
\end{equation}
and $u(s)$ satisfies 
\begin{equation}
    bs+\left(bs e^s-1\right)u-u^2 e^s-bs e^{u\eta_c}=0\ .\label{eqforu:main}
\end{equation}

\begin{comment}
\begin{figure}[t]
    \centering
    \includegraphics[scale = 0.23]{picturejump2.pdf}
    \caption{{\bf Main panel:} sketch of a trajectory with positive jumps that crosses $L$ at step $n$. {\bf Inset (top):} behavior of the lower edge $\ell_1(b)$ of the support of the rate function $\Phi(z)$ as a function of $b$. {\bf Inset (bottom):} critical line in the $(b,\eta_c)$ plane separating the regions $[A]$ and $[B]$ where the rate function $\Phi(z)$ attains a finite value at the lower edge $\ell_1(b)$ or diverges logarithmically, respectively. }
    \label{fig:jump}
\end{figure}
\end{comment}

The rate function $\Phi(z)$ is supported over $z\in [\ell_1(b),\infty)$ and has the following asymptotic behaviors \cite{supplemental}
\begin{equation}
   \Phi(z)\approx 
   \begin{cases}
   z\ln z-z+1 &\qquad z\to\infty\\
   \frac{(z-a_1)^2}{2\sigma_C^2} &\qquad z\sim a_1\\
   \psi_b(z) &\qquad z\to \ell_1(b)\,,
   \end{cases} \label{asymp}
\end{equation}
where $\ell_1(b)$ and $\psi_b(z)$ are given below (see Eq.~\eqref{eq:psib}).
Thus, in the typical regime, the cost fluctuates around the typical value $C=a_1 L$, where $a_1=1+b \exp(-\eta_c)$, with variance $\mathrm{Var}(C|L)=L\sigma_C^2$, where
\begin{equation}
    \sigma_C^2 = 1+2b^2 e^{-\eta_c}(1-e^{-\eta_c})-2b\eta_c e^{-\eta_c}(1+b e^{-\eta_c})\ .\label{varianceFPcost}
\end{equation}

\begin{figure}[ht]
    \centering
   \includegraphics[scale = 0.55]{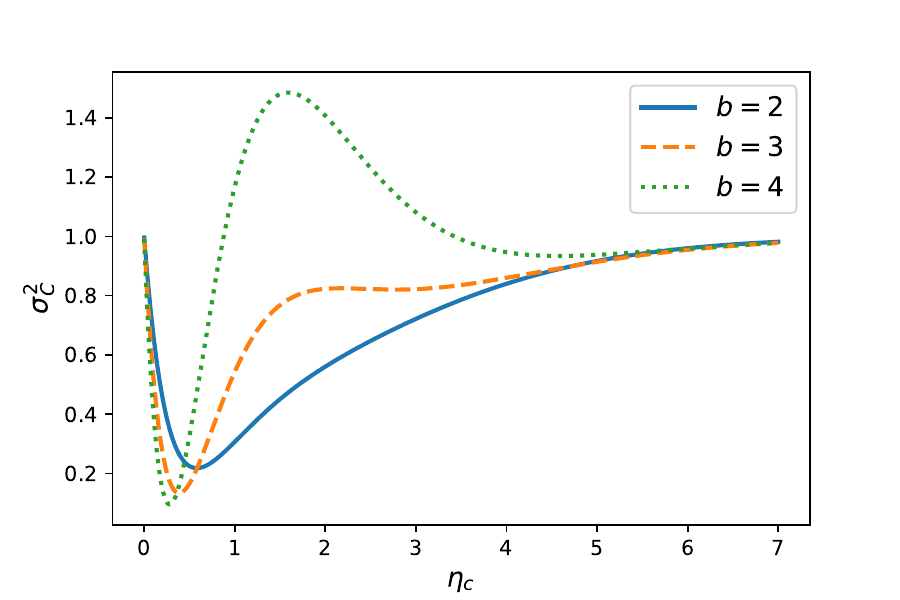}
    \caption{Variance $\sigma_C^2$ of the hitting cost as a function of $\eta_c$ for increasing values of $b$.}
    \label{fig:variance}
\end{figure}

It turns out that the cost variance $\sigma_C^2$ has a surprisingly rich behavior as a function of the parameters $\eta_c$ and $b$. First, for both $\eta_c\to 0$ and $\eta_c\to\infty$ the scaled variance $\sigma_C^2$ tends to the limiting value $1$ (see Fig. \ref{fig:variance}). This counter-intuitive result can be understood as follows. For $\eta_c\to\infty$, all of the steps are time-like and hence the cost $C= n$, where $n$ is the number of steps. The distribution $P(n|L)$ of the number of steps $n$ needed -- given the target location $L$ -- is Poisson, $P(n|L)=e^{-L}L^{n-1}/(n-1)!$ for $n=1,2,\ldots$ (see \cite{supplemental} for the derivation). Therefore, $\operatorname{Var}(C|L)=\operatorname{Var}(n|L)= L$. On the other hand, for $\eta_c\to 0$, all of the steps are space-like and hence $C=n+bL$, where $L$ is the final position. Thus, we obtain again $\operatorname{Var}(C|L)\approx \operatorname{Var}(n|L)=L$. 

For intermediate values of $\eta_c$, the behavior of $\sigma_C^2$ depends on $b$. For small values of $b$, $\sigma_C^2$ has a unique minimum as a function of $\eta$ (see \cite{supplemental}). Interestingly, above the critical value $b>b_c\approx 2.953\ldots$ (that we computed numerically with Mathematica), $\sigma_C^2$ develops a second minimum and a maximum (see Fig.~\ref{fig:variance}). This behavior can be qualitatively understood as follows: for $\eta_c$ slightly above zero, most of the step are still space-like. Therefore, $C= \sum_{k=1}^n h(\eta_k)\approx b L +(1- b\eta_c) n$, hence the variance $\operatorname{Var}(C|L)\approx (1- b\eta_c)^2 L$, implying that $\sigma_C^2 = (1-b\eta_c)^2$ initially must decrease linearly as $\eta_c$ increases from zero. As $\eta_c$ increases further, `time-like' steps become more and more abundant. The cost fluctuations start increasing again with increasing $\eta_c$ and become maximal at some $\eta_c^\star$. These `giant' fluctuations reflect the perfect mixing of time-like and space-like steps, which can be arranged in the maximal number of different ways to cover the distance $L$. Increasing $\eta_c$ further beyond the maximum leads the cost fluctuations to subside, as the pure ``time-like'' phase settles in.

\vspace{-8pt}

\begin{figure}[ht]
    \centering
   \includegraphics[scale = 0.55]{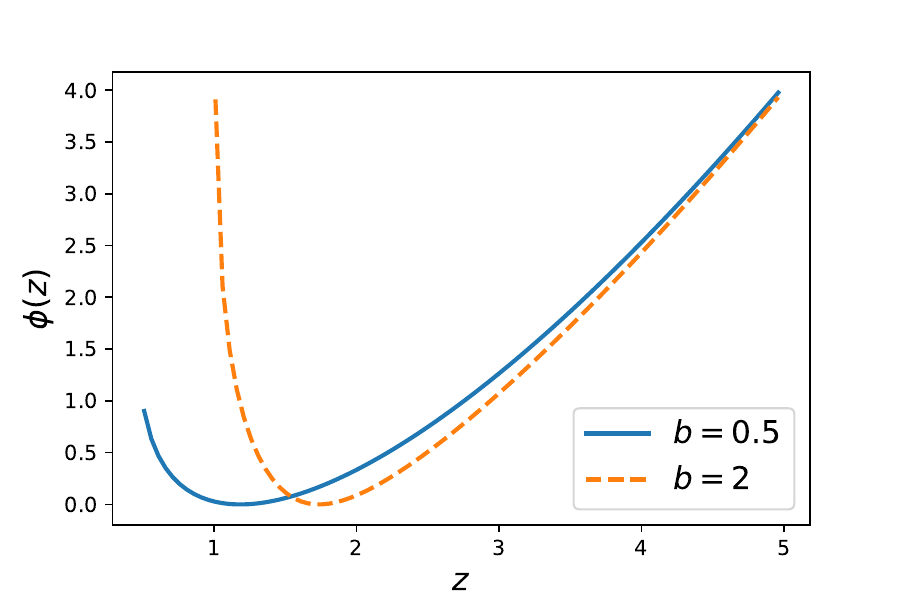}
    \caption{Rate function $\Phi(z)$ describing the large deviations of the hitting cost, evaluated by solving Eq. \eqref{legendrephi} and Eq.~\eqref{eqforu:main} numerically with Mathematica for $\eta_c=1$ and $b=0.5$ (continuous line) and $\eta_c=1$ and $b=2$ (dashed line).}
    \label{fig:rate_functions}
\end{figure}

The behavior of the lowest edge $\ell_1(b)$ of the support of $\Phi(z)$ is also very interesting. First, the edge $\ell_1(b)$ itself depends on whether $b$ is smaller or larger than $1/\eta_c$. More precisely, $\ell_1(b)=b$ for $b\leq 1/\eta_c$, while $\ell_1(b)=1/\eta_c$ for $b\geq 1/\eta_c$.
Around the lower edge, we have the following behavior for $z\to \ell_1(b)^+$
\begin{equation}
    \psi_b(z)=
    \begin{cases}
        1+\frac{z-b}{1-b\eta_c}\ln\left(\frac{z-b}{1-b\eta_c}\right)-\frac{z-b}{1-b\eta_c} &\qquad b<1/\eta_c\\
        -2\sqrt{\delta_1}\sqrt{z-b}+1+\delta_0&\qquad b=1/\eta_c\\
        -\frac{1}{\eta_c}\ln\left(z-\frac{1}{\eta_c}\right)+\mathcal{O}(1)&\qquad b>1/\eta_c\\
    \end{cases}
    \label{eq:psib}
\end{equation}
with $\delta_0$ solution of $\delta_0+e^{\delta_0/b}=0$ and $\delta_1=\delta_0^2/(b-\delta_0)$. Therefore, the rate function $\Phi(z)$ attains a finite value at the lower edge of its support, $\ell_1(b)=b$, if $b\leq 1/\eta_c$, whereas it diverges logarithmically at the lower edge of its support, $\ell_1(b)=1/\eta_c$, if $b> 1/\eta_c$. A plot of the full rate function computed by solving Eq. \eqref{legendrephi} and Eq.~\eqref{eqforu:main} numerically with Mathematica, along with the asymptotic behaviors above is included in Fig.~\ref{fig:rate_functions} for different values of $b$ and $\eta_c=1$.

Interestingly, the lower edge $\ell_1(b)$ ``freezes'' to the value $\ell_1(b)=1/\eta_c$ for $b>1/\eta_c$. To understand this freezing transition, we notice that the lower edge $\ell_1(b)$ is related to the minimal possible cost $C_{\rm min}$ by $\ell_1(b)=C_{\min}/L$. In \cite{supplemental}, we show that if $b<1/\eta_c$, i.e., when space-like configurations are sufficiently inexpensive, the cost is minimized by a single long jump of length $\eta=L$, corresponding to $C_{\min}\approx bL$ (and hence $\ell_1(b)=b$). On the other hand, for $b>1/\eta_c$, the minimal cost $C_{\min}=n=L/\eta_c$ is attained with $n=L/\eta_c$ time-like steps of length $\eta_c$ leading to  $\ell_1(b)=1/\eta_c$.

\textit{Derivations - }We focus on the probability $P(C,X|n)$ that a random walker that has reached position $X$ after exactly $n$ steps with a total cost $C$ (Ensemble (i)). This quantity can be formally written as
\begin{equation}
 P(C,X|n)=\left\langle \delta\left(X-\sum_{k=1}^n \eta_k\right)\delta\left(C-\sum_{k=1}^n h(\eta_k)\right)\right\rangle\ ,  
 \label{eq:G_def}
\end{equation}
where the average is performed over the variables $\eta_i$. Taking the double Laplace transform $\tilde P (\lambda, s|n)=\int_0^\infty  P(C,X|n)e^{-\lambda X-sC}dXdC$ yields \cite{supplemental} $\tilde P (\lambda, s|n)=[g(\lambda,s)]^n$, where
\begin{equation}
    g(\lambda,s)=\frac{e^{-s}}{\lambda+1}\left[1-\frac{bs}{\lambda+1+bs}e^{-(\lambda+1)\eta_c}\right]\ .\label{glambdas}
\end{equation}
The distribution of the final position alone is easily obtained as $P(X|n)=(1/\Gamma(n))X^{n-1}e^{-X}$ for an exponential jump distribution. Thus, the $k$-th moment of $C$, conditioned on the total displacement $X$, can be obtained by Laplace inversion \cite{supplemental}, leading to the exact expressions for the average and variance of the cost in Eqs.~\eqref{mean_en2_exact} and \eqref{var_rel.1} of \cite{supplemental}. The fact that the conditional variance is nonzero embodies the \emph{taxi paradox} described earlier, as different trajectories reaching the same spot after the same number of steps ($=$ time) may indeed charge different amounts!

\begin{comment}
We then briefly outline how to derive the large deviation form in \eqref{eq:large_deviation} for Ensemble (ii). The joint distribution of the jumps needed to reach $L$, the number of such jumps (which is no longer fixed, but random), and the total cost is
\begin{equation}
P(\{\eta_i\},n,C|L)={\cal P}(\vec \eta) \delta\left(\sum_{i=1}^{n}\eta_i-L\right)\delta\left(\sum_{i=1}^{n} h(\eta_i)-C\right)\ ,\label{jointPPpaper}
\end{equation}
where ${\cal P}(\vec \eta) =e^{-\sum_{i=1}^{n} \eta_i}$. Taking the Laplace transform $\int_0^\infty dC dL~\exp(-sC-\lambda L)P(\{\eta_i\},n,C|L)$ and marginalizing over $\{\eta_i\}$ and $n$, we get \cite{supplemental}
\begin{equation}
 \hat P(s,\lambda) =\frac{1}{1-g(\lambda,s)}\ ,
\end{equation}
with $g(\lambda,s)$ defined in \eqref{glambdas}. Taking the inverse Laplace transform over $\lambda$ only, we get for the Laplace transform of the hitting cost distribution $P(C|L)$
\begin{equation}
    \hat P(s|L)=\int_\Gamma \frac{d\lambda}{2\pi\mathrm{i}}\frac{1}{1-g(\lambda,s)}e^{\lambda L}\ ,
\end{equation}
where $\Gamma$ is a Bromwich contour. To analyze the behavior of this integral for large $L$, it is necessary to consider the value $\lambda^\star$ for which the integrand has a pole, namely $g(\lambda^\star,s)=1$. This leads to 
$
\hat P(s|L)\sim \exp(L \lambda^\star (s))$
for large $L$. Setting $\lambda^\star(s)+1=-u(s)$, the large-$L$ behavior in Eq. \eqref{eq:large_deviation} follows from the standard Legendre-Fenchel inversion \cite{supplemental}, with rate function given in \eqref{legendrephi}. 
\end{comment}

\textit{Conclusions and Outlook - } Motivated by the taxi paradox,
we have introduced and solved exactly a simple model for diffusion with a nonlinear cost associated to each 
jump. Our results exhibit unexpected phenomena, including giant fluctuations of the 
cost and a freezing transition in the large deviation regime of the total cost. We expect that our results should 
apply generally to arbitrary jump 
distributions $p(\eta)$ with a finite variance. We have shown that our results can be directly applied
to a variety of physical systems where an extended object is pulled by a random force in
a disordered medium.
In future works, it would be interesting to investigate fat-tailed jump 
distributions such as L\'evy walks, which are of central importance in 
finance and biology~\cite{ZDK2015}. In particular, very fat-tailed jump 
distribution will display condensation phenomena, where a single jump dominates the trajectory \cite{Majumdar2005}. It 
would be relevant to investigate the impact of a nonlinear cost on such setting. Moreover, one may consider cost 
functions that penalize short jumps and rewards instead long excursions with a flat fee -- a pattern commonly found in 
public transportation pricing models, where monthly passes are typically cheaper than collecting single ride tickets.

\textit{Acknowledgements - }This work was supported by a Leverhulme Trust International Professorship grant [number 
LIP-202-014]. For the purpose of Open Access, the authors have applied a CC BY public copyright licence to any Author 
Accepted Manuscript version arising from this submission.

\newpage
\onecolumngrid

\section{Supplemental Material}

Consider a random walk (jump process) in discrete time and continuous space, where each jump
length is a positive independent and identically distributed (IID) variable drawn from a continuous PDF $p(\eta)$ (normalized to unity).
Let us denote the cumulative distribution
\begin{equation}
q(\eta)= \int_{\eta}^{\infty} p(\eta')\, d\eta' \, .
\label{cumul.1}
\end{equation}
Let $X$ and $C$ denote the position and cost after $n$ steps, i.e.,
\begin{equation}
X= \sum_{i=1}^n \eta_i \, \quad {\rm and}\quad C= \sum_{i=1}^n h(\eta_i)\ ,
\label{position_cost}
\end{equation}
where $h(\eta)= 1+ b (\eta-\eta_c)\, \theta(\eta-\eta_c)$ is the non-linear cost associated to each jump.
Below, we consider two different ensembles where the underlying random variables are different, and then demonstrate how
the statistics of total cost in the two models are related to each other.

\subsection{Ensemble (i)}

In this Ensemble, the cost $C$ and the position $X$ are random variables, but
the number of steps $n$ is fixed. Let $P(C,X|n)$ denote the joint distribution of $C$ and $X$, given fixed $n$.
It is useful to write this joint distribution explicitly by integrating over the underlying jump lengths. 
One obtains
\begin{equation}
P(C,X|n)= \int d\eta_1\int d\eta_2 \cdots \int d\eta_n\, p(\eta_1)p(\eta_2)\cdots p(\eta_n)\, 
\delta\left(\sum_{i=1}^n \eta_i-X\right)\, \delta\left(\sum_{i=1}^n h(\eta_i)-C\right) \ ,
\label{en1.1}
\end{equation}
where all the integrals (here and below) run between $0$ and $\infty$. We emphasize that in this ensemble $C$ and $X$ are random variables and $n$ is fixed. This means that
when we integrate over $C$ and $X$ for any $n$, we should get $1$. This is easily verified from
Eq. \eqref{en1.1}, since $p(\eta_i)$ is normalized to unity for each $\eta_i$.
By integrating over $C$, we can easily compute the marginal distribution $P(X|n)$ as
\begin{equation}
P(X|n) = \int_{0}^{\infty} dC\, P(C,X|n) = \int d\eta_1\int d\eta_2 \ldots \int d\eta_n \, p(\eta_1)p(\eta_2)\ldots p(\eta_n)\,
\delta\left(\sum_{i=1}^n \eta_i-X\right)\,\, .
\label{en2.1}
\end{equation}
For example, for positive exponential distribution, $p(\eta)= e^{-\eta}\, \theta(\eta)$, it is easy to see
that $P(X|n)$ is given by (taking Laplace transform of Eq. (\ref{en2.1}) with respect to $X$ and inverting)
\begin{equation}
P(X|n)= e^{-X}\, \frac{X^{n-1}}{\Gamma(n)}\, ,
\label{en1_px.1}
\end{equation}
which is normalized to unity, $\int_0^{\infty} P(X|n)\, dX=1$. Knowing this marginal distribution,
one can compute the conditional distribution of $C$, given $X$ and $n$ as
\begin{equation}
P_{\rm cond}(C|X,n)= \frac{P(C,X|n)}{P(X|n)}\, .
\label{en1_cond.1}
\end{equation}
One then computes the mean and the variance of $C$ from this conditional distribution for fixed $X$ and $n$, i.e,.
\begin{align}
\langle C\rangle_{X,n} &=   \int_0^{\infty} dC\, C\, P_{\rm cond}(C|X,n)= \int_0^{\infty} dC\, C\,
\frac{P(C,X|n)}{P(X|n)}= \frac{1}{P(X|n)}\, \int_0^{\infty} dC\, C\, P(C,X|n)\ , \label{mean_en1} \\ 
{\rm Var}(C|X,n) &= \frac{1}{P(X|n)}\, \int_0^{\infty} dC\, C^2\, P(C,X|n) 
- \left[\langle C\rangle_{X,n}\right]^2\, .
\label{var_en1.1}
\end{align}

These quantities will be computed below.

\subsection{Ensemble (ii)}

In this second ensemble, the random variables are $C$ and $n$, but the final position is kept fixed at $X=L$. Let $P(C,n|L)$ denote the joint distribution of $C$ and $n$, for fixed $L$.
One can again express this joint distribution by integrating over the underlying random jumps as
\begin{equation}
P(C,n|L)= \int d\eta_1\int d\eta_2 \cdots \int d\eta_n\, p(\eta_1)p(\eta_2)\cdots q(\eta_n)\,
\delta\left(\sum_{i=1}^n \eta_i-L\right)\, \delta\left(\sum_{i=1}^n h(\eta_i)-C\right) \, ,
\label{jpdf_en2.1}
\end{equation}
where $q(\eta)$ is the cumulative distribution defined in Eq. \eqref{cumul.1}. Note that for a generic
$p(\eta)$, the two distributions respectively in Eqs. \eqref{en1.1} and \eqref{jpdf_en2.1} differ
from each other only in the last jump $q(\eta_n)$. Eq. \eqref{jpdf_en2.1} can be understood as follows.
Once the walker has made $(n-1)$ complete jumps, it has to reach $L$ in the last jump. So, the
length of the last jump has a different distribution from the preceding ones due to the constraint of reaching $L$. In fact, the statistical weight attached to this last jump
is  $q(\eta_n)= \int_{\eta_n}^{\infty} p(\eta)\, d\eta$ since the last jump would have been `completed' after $\eta_n$. 
This is a standard computation in renewal processes. The important thing is to check whether it satisfies
the correct normalization. To check this normalization, we first integrate Eq. \eqref{jpdf_en2.1}
over $C$ to get the marginal PDF $P(n|L)$ as
\begin{equation}
P(n|L)= \int_{0}^{\infty} P(C,n|L)\, dC= \int d\eta_1\int d\eta_2 \cdots \int d\eta_n\, p(\eta_1)p(\eta_2)\cdots q(\eta_n)\,
\delta\left(\sum_{i=1}^n \eta_i-L\right)\ .
\label{nmarg_en2.1}
\end{equation}
Let us now take the Laplace transform with respect to $L$ to write
\begin{equation}
\int_0^{\infty} P(n|L)\, e^{-\lambda\, L}\, dL= \left[\tilde{p}(\lambda)\right]^{n-1}\, \tilde{q}(\lambda)\,,
\label{norm_check.1}
\end{equation}
where $\tilde{p}(\lambda)=\int_0^{\infty} p(\eta)\, e^{-\lambda\, \eta} \, d\eta$ is the Laplace
transform of the jump distribution. 
It is easy to see, using integration by parts, that
\begin{equation}
\tilde{q}(\lambda)= \int_0^{\infty} q(\eta)\, e^{-\lambda\, \eta}\, d\eta= \int_0^{\infty}
\left(\int_{\eta}^{\infty} p(\eta')\, d\eta' \right)\, e^{-\lambda\, \eta}\, d\eta= 
\frac{1-\tilde{p}(\lambda)}{\lambda}\ .
\label{norm_check.2}
\end{equation}
Substituting this result in Eq. \eqref{norm_check.1} gives
\begin{equation}
\int_0^{\infty} P(n|L)\, e^{-\lambda\, L}\, dL=
\left[\tilde{p}(\lambda)\right]^{n-1}\, \frac{1-\tilde{p}(\lambda)}{\lambda}\, .
\label{norm_check.3}
\end{equation}
Finally, if we sum Eq. \eqref{norm_check.3} over $n=1,2,\ldots$  we get
\begin{equation}
\int_0^{\infty} \sum_{n=1}^{\infty} P(n|L)\, e^{-\lambda\, L}\, dL= \frac{1}{\lambda}\ .
\label{norm_check.4}
\end{equation}
Inverting with respect to $\lambda$, we get our desired normalization
\begin{equation}
\sum_{n=1}^{\infty} P(n|L) = 1\, \quad {\rm for}\,\, {\rm all}\,\, L \, .
\label{norm.5}
\end{equation}

Thus, Eq. \eqref{jpdf_en2.1} is an appropriately normalized joint distribution for the two random variables
$C$ and $n$, with fixed $L$. The marginal distribution of $C$ for fixed $L$ is then obtained
by summing over the random variable $n$
\begin{equation}
P(C|L)= \sum_{n=1}^{\infty} P(C,n|L)\, .
\label{margc_en2.1}
\end{equation}
Finally, we compute the mean cost and its variance for fixed $X$ in Ensemble (ii) as
\begin{align}
\langle C \rangle_{L} &= \int_0^{\infty} dC\, C\, P(C|L)= 
\sum_{n=1}^{\infty} \int_0^{\infty} dC\, C\, P(C,n|L) \label{mean2.1} \\
{\rm Var}(C|L) &= \int_0^{\infty} dC\, C^2\, P(C|L)-\left[\langle C \rangle_{L}\right]^2
= \sum_{n=1}^{\infty} \int_0^{\infty} dC\, C^2\, P(C,n|L)-\left[\langle C \rangle_{L}\right]^2\, .
\label{var2.1}
\end{align} 

\subsection{Relationship between the two ensembles}

In this section, we will be using the notation $X$ (instead of $L$) for Ensemble (ii) to indicate the fixed total distance in order to compare the two ensembles. The two JPDF's, namely $P(C,X|n)$ in Eq. \eqref{en1.1} and $P(C,n|X)$ in
Eq. \eqref{jpdf_en2.1}, are different from each other for generic $p(\eta)$. This is natural
because the random variables in the two ensembles are different. Hence, for generic $p(\eta)$,
there is no simple relation between the moments of the total cost $\langle C^k\rangle$ in the two Ensembles. 
In particular, the
two variances ${\rm Var}(C|X,n)$ in Eq. \eqref{var_en1.1} and ${\rm Var}(C|X)$ in Eq. \eqref{var2.1} have
no reasons to be related to each other.
However,
for the special case of exponential jump distribution $p(\eta)= e^{-\eta}$, it is clear that
$q(\eta)= \int_{\eta}^{\infty} p(\eta')\, d\eta'= e^{-\eta}$ and hence, in this case, Eqs. \eqref{en1.1}
and \eqref{jpdf_en2.1} coincide exactly and we have
\begin{equation}
P(C,X|n)= P(C,n|X)= \int d\eta_1\int d\eta_2 \cdots \int d\eta_n \, e^{-\sum_{i=1}^N \eta_i} \,
\delta\left(\sum_{i=1}^n \eta_i-X\right)\, \delta\left(\sum_{i=1}^n h(\eta_i)-C\right) \ .
\label{equiv.1}
\end{equation}
Thus for the special exponential distribution $p(\eta)=e^{-\eta}$, the two joint
distributions $P(C,X|n)$ and $P(C,n|X)$ coincide,
although the underlying random variables in the two ensembles are quite different.
To see how the cost moments $\langle C^k\rangle$ in the two Ensembles may be related in this special case, we first compute
the marginal distribution $P(n|X)= \int_0^{\infty} P(C,n|X)\, dC$ in Ensemble (ii).
Substituting $\tilde{p}(\lambda)=1/(1+\lambda)$ and using $\tilde{q}(\lambda)=\tilde{p}(\lambda)$ in Eq. \eqref{norm_check.3} we get
\begin{equation}
\int_0^{\infty} P(n|X)\, e^{-\lambda\, X}\, dX= \frac{1}{(1+\lambda)^n}\, ,
\label{margn_2}
\end{equation}
which, upon inversion, gives the Poisson distribution for $n\ge 1$
\begin{equation}
P(n|X)= e^{-X}\, \frac{X^{n-1}}{\Gamma(n)}\, .
\label{poisson.1}
\end{equation}
By comparing this to Eq. \eqref{en1_px.1}, we see that 
\begin{equation}
P(n|X)=P(X|n)\, .
\label{marg_equiv}
\end{equation}
However, once again, one should remember
that the
underlying random variable is $n=1,2,\ldots$ in $P(n|X)$ , while it is the
continuous variable $X$ in $P(X|n)$. Using the two identities 
$P(C,X|n)= P(C,n|X)$ and $P(X|n)=P(n|X)$ for this exponential jump distribution, it then follows from Eqs. \eqref{mean_en1}
and \eqref{mean2.1} that
\begin{equation}
\langle C \rangle_{X}= \sum_{n=1}^{\infty} \int_0^\infty dC\, C\, P(C,n|X)
=\sum_{n=1}^{\infty} \int_0^{\infty} dC\, C\, P(C,X|n)
=\sum_{n=1}^{\infty} P(X|n) \, \langle C\rangle_{X,n}
=\sum_{n=1}^{\infty} P(n|X) \, \langle C\rangle_{X,n} \, .
\label{mean_rel.1}
\end{equation}
Using $P(X|n)=P(n|X)=e^{-X}\, \frac{X^{n-1}}{\Gamma(n)}$, then gives the precise relation between the mean costs
in the two Ensembles
\begin{equation}
\langle C \rangle_{X} = \sum_{n=1}^{\infty} e^{-X}\, \frac{X^{n-1}}{\Gamma(n)}\, \langle C\rangle_{X,n}\, .
\label{mean_rel.2}
\end{equation}
As a useful check of this exact relation, note that the average total cost in Ensemble (i) is (see Eq. (13) in the main text, and derivation below)
\begin{equation}
\langle C\rangle_{X,n}= n + b\, X\, \left(1-\frac{\eta_c}{X}\right)^{n}\, \theta(X-\eta_c)\, ,
\label{mean_en1_exact}
\end{equation}
valid for all $n\ge 1$ and all $X$. Substituting this result on the right hand side
of Eq. \eqref{mean_rel.2} and summing over $n$ exactly, we get
\begin{equation}
\langle C\rangle_{X}= 1+X+ b\, (X-\eta_c)\, e^{-\eta_c}\, \theta(X-\eta_c)\, .
\label{mean_en2_exact}
\end{equation}
Now, for large $X$, this leads to
\begin{equation}
\langle C\rangle_{X} \approx X\, \left(1+ b\, e^{-\eta_c}\right)\, ,
\label{mean_en2_largeX}
\end{equation}
which coincides exactly (replacing $X$ by $L$) with $L \, a_1= L(1+ b~ e^{-\eta_c})$ as stated
just before Eq.~\eqref{varianceFPcost} in the main text. This is a useful check.

In an identical way, the $k$-th moment of the cost in the two ensembles are also related via
\begin{equation}
\langle C^k \rangle_{X} = \sum_{n=1}^{\infty} e^{-X}\, \frac{X^{n-1}}{\Gamma(n)}\, \langle C^k\rangle_{X,n}\, .
\label{moment_rel.2}
\end{equation}
Consequently, the variances in the two ensembles are also related, but the relationship is not as simple
as the moments in Eq. \eqref{moment_rel.2}. One has to compute the first and the second moment in each ensembles
(which are related via Eq. \eqref{moment_rel.2}), and then compute $\langle C^2\rangle- \langle C\rangle^2$ for
the variance. In other words, the precise relation between the two variances is
\begin{equation}
{\rm Var}(C|X)+ \left[\langle C\rangle_X\right]^2 = \sum_{n=1}^{\infty} e^{-X}\, \frac{X^{n-1}}{\Gamma(n)}\, 
\left[ {\rm Var}(C|X,n)+   \left(\langle C\rangle_{X,n}\right)^2\right]\, .
\label{var_rel.1}
\end{equation}

\subsection{Ensemble (i): Calculation of the conditional average and variance of the total cost\label{sec}}

In this Section, we compute the average $\langle C\rangle_{X,n}$ and the variance ${\rm Var}(C|X,n)$ of the total cost charged in Ensemble (i) for an excursion that reaches $X$ in exactly $n$ steps. We consider again the exponential jump pdf $p(\eta)=e^{-\eta}$.

We start from the probability $P(C,X|n)$ (given $n$) in Eq.~\eqref{en1.1} that reads
\begin{equation}
 P(C,X|n)=\left\langle \delta\left(X-\sum_{k=1}^n \eta_k\right)\delta\left(C-\sum_{k=1}^n h(\eta_k)\right)\right\rangle\ ,
\end{equation}
where $\langle \ldots\rangle$ denotes the average over the jump variables $\eta_i$'s. We next take the Laplace transform with respect to $X$ and $C$
\begin{equation}
\tilde P (\lambda, s|n)=\int_0^\infty dC \int_0^\infty dX e^{-s C- \lambda X} P(C,X|n)= [g(\lambda,s)]^n 
\end{equation}
where 
\begin{equation}
    g(\lambda, s)=\int_0^\infty d\eta~p(\eta)e^{-\lambda \eta-s h(\eta)}= \frac{e^{-s}}{\lambda+1}\left[1-\frac{bs}{\lambda+1+bs}e^{-(\lambda+1)\eta_c}\right]\ .
\end{equation}
Here, we used the fact that in Laplace space, all $\eta$-integrals decouple, and are identical to each other.

Then, taking the $k$-th
derivative of $\tilde P (\lambda, s|n)$ with respect to $s$ and setting $s=0$, 
one gets the
Laplace transform (with respect to $X$) of the $k$-th moment of the cost
\begin{equation}
\int_0^{\infty} \int_0^{\infty} dC~ C^k P(C,X|n) e^{-\lambda X} dX
= (-1)^k \frac{d^k}{d s^k} [g(\lambda,s)]^n \Big|_{s=0}\ .\label{SM:derivatives}
\end{equation}

For example, for $k=1$ and $k=2$ we get after simple algebra from \eqref{SM:derivatives} the two expressions \begin{align}
\int_0^{\infty} \int_0^{\infty} dC C P(C,X|n) e^{-\lambda X} dX
&=n \left(\frac{1}{\lambda +1}\right)^{n+1} \left(b e^{-\eta_c  (\lambda +1)}+\lambda +1\right)\label{SMinvfirst}\\
\int_0^{\infty} \int_0^{\infty} dC C^2 P(C,X|n) e^{-\lambda X} dX
&=n e^{-2 \eta_c  (\lambda +1)} \left(\frac{1}{\lambda +1}\right)^{n+2} \nonumber \\&\times  \left(b^2 (n-1)+2 b e^{\eta_c  (\lambda +1)} (b+\lambda  n+n)+(\lambda +1)^2 n e^{2 \eta_c  (\lambda +1)}\right)\ .\label{SMinvsecond}
\end{align}
Then, we can invert the Laplace transforms \eqref{SMinvfirst} and \eqref{SMinvsecond} with respect to $\lambda$ using
\begin{align}
\label{eq:Laplace_inversions}
\mathcal{L}^{-1}_\lambda\left[\frac{e^{-A\lambda}}{(\lambda+1)^B}\right](X) &=\frac{(X-A)^{B-1} \theta (X-A) e^{-(X-A)}}{\Gamma (B)}\,,\\
\mathcal{L}^{-1}_\lambda\left[\frac{\lambda e^{-A\lambda}}{(\lambda+1)^B}\right](X) &=-\frac{(X-A)^{B-2} \theta (X-A) e^{-(X-A)} (-A-B  +1 +X)}{ \Gamma (B)}\,.
\end{align}
Using these Laplace-inversion formulae, we invert Eqs.~\eqref{SMinvfirst} and \eqref{SMinvsecond}
to compute $\int_0^{\infty} dC~ C P(C,X|n)$ and $\int_0^{\infty} dC~ C^2 P(C,X|n)$
explicitly. Next we divide by the marginal distribution \begin{equation}
\label{eq:PXNsm}
    P(X|n)= e^{-X} X^{n-1}/(n-1)!
\end{equation}
to compute $\langle C\rangle_{X,n}= \int_0^{\infty} dC~ C P(C,X|n)/P(X|n)$
and $\langle C^2\rangle_{X,n}= \int_0^{\infty} dC~ C^2 P(C,X|n)/P(X|n)$.
After lengthy but straightforward algebra, this leads to the following exact results for the conditional mean and the variance 
 \begin{align}
&\langle C\rangle_{X,n} =b X \theta (X-\eta_c ) \left(1-\frac{\eta_c }{X}\right)^n+n\ ,\label{conditionalexpectation}\\ 
\nonumber &\mathrm{Var}(C|X,n) =b^2 X^2\left\{\left[\frac{n-1}{n+1}\left(1-2\frac{\eta_c}{X}\right)^{n+1}\right]\theta(X-2\eta_c)+\right.\\
&\left. +\left[\frac{2}{(n+1)}\left(1-\frac{\eta_c}{X}\right)^{n+1}-\left(1-\frac{\eta_c}{X}\right)^{2n}\right]\theta(X-\eta_c)\right\}\label{condvar}\ ,
\end{align}
as given in the main text. Taking the scaling limit $X\to\infty$, $n\to\infty$, with $X/(n\eta_c)=y$ fixed, we obtain
\begin{align}
    \langle C\rangle_{X,n}\to n+ b\eta_c n H\left(\frac{X}{n\eta_c}\right)\\
     \operatorname{Var}(C|X,n)\to b^2 \eta_c^2 n F\left(\frac{X}{n\eta_c}\right)\ ,
\end{align}
with
\begin{align}
H(y) &=y e^{-1/y}\ ,\\
    F(y) &= e^{-2/y} \left(2 e^{1/y} y^2-2 y^2-2 y-1\right)\ ,
    \label{Fysm}
\end{align}
as given in the main text.

\subsection{Ensemble (ii): Poisson distribution for the number of steps needed, given a target location at $L$}

We start from the joint distribution of the jumps and the number of steps, given the fixed final location $X=L$
\begin{equation}
    P(\{\eta_i\},n|L)=e^{-\sum_{i=1}^n \eta_i}\delta\left(\sum_{i=1}^n \eta_i - L\right)\ .
\end{equation}
Taking the Laplace transform w.r.t. $L$, and integrating over the jumps, we get for the Laplace transform of the marginal distribution $P(n|L)$ of the number of steps given the target location at $L$
\begin{align}
\nonumber \hat P(n;\lambda) &=\int_0^\infty dL~e^{-\lambda L}\int_0^\infty d\eta_1 \cdots \int_0^\infty d\eta_n P(\{\eta_i\},n|L)\\
&= \left[\int_0^\infty d\eta~e^{-(1+\lambda)\eta}\right]^n=\frac{1}{(1+\lambda)^n}\ .
\end{align}
Inverting the Laplace transform, we get the Poisson distribution with parameter $L$
\begin{equation}
    P(n|L)=e^{-L}\frac{L^{n-1}}{(n-1)!}
\end{equation}
for $n=1,2,\ldots$ as claimed in the main text.

\subsection{Ensemble (ii): Asymptotics of $\Phi(z)$}

Consider that our random process stops when the target location $L$ is reached. Let $\{\eta_1,\ldots,\eta_n\}$ denote the jump lengths in a typical configuration, where of course the number of steps $n$ to reach $L$ is now a random variable.

The joint distribution of the jump lengths, the number of steps $n$ and the total cost $C$ -- given the target position $L$ is given by 

\begin{comment}
\begin{equation}
P(\eta_1,\ldots,\eta_{n-1},\eta_n',n,C|L)=e^{-\sum_{i=1}^{n-1} \eta_i}f(\eta_n')\delta\left(\sum_{i=1}^{n-1}\eta_i+\eta_n'-L\right)\delta\left(\sum_{i=1}^{n-1} h(\eta_i)+h(\eta_n')-C\right)\ ,\label{jointPP}
\end{equation}
where the first $n-1$ steps are drawn from the (exponential) jump distribution, and the last ``incomplete'' step $\eta_n$ -- which is just enough to reach the target $L$ -- is drawn from the pdf $f(\eta')$, which is easily computed from the following reasoning. 

In the absence of a target location (which kills the process as soon as it is reached), the $n$-th step $\eta_n$ would be exponentially distributed. This $n$-th step $\eta_n$ can be seen as the sum of two contributions, $\eta_n = \eta_n' + \eta_n''$, where the first mini-step $\eta'$ is long just enough to take the walker to $L$, and the second mini-step $\eta''$ can be as long as needed to make up the ``full'' step $\eta_n$. Therefore
\begin{equation}
    f(\eta')= \int_0^\infty d\eta'' \int_0^\infty d\eta~\delta(\eta- (\eta'+\eta''))e^{-\eta}=e^{-\eta'}\ .
\end{equation}
Therefore, the last incomplete step is distributed according to all the previous others, and we can simplify the notation in Eq. \eqref{jointPP} as
\end{comment}

\begin{equation}
P(\{\eta_i\},n,C|L)=e^{-\sum_{i=1}^{n} \eta_i}\delta\left(\sum_{i=1}^{n}\eta_i-L\right)\delta\left(\sum_{i=1}^{n} h(\eta_i)-C\right)\ .\label{jointPP2}
\end{equation}

Marginalizing over $\{\eta_i\}$ and $n$, and taking the Laplace transform with respect to $L$ and $C$, we get
\begin{align}
  \nonumber  \hat P(s,\lambda) &=\int_0^\infty dC~e^{-sC}\int_0^\infty dL~e^{-\lambda L}\sum_{n=1}^\infty \int_0^\infty\cdots \int_0^\infty d\eta_1\cdots d\eta_n P(\{\eta_i\},n,C|L)\\
  &=\sum_{n=1}^\infty \left[\int_0^\infty d\eta~e^{-(1+\lambda)\eta-s h(\eta)}\right]^n=\frac{1}{1-g(\lambda,s)}\ ,
\end{align}
where we used the standard geometric series, and $g(\lambda,s)$ is defined in Eq. (12) of the main text.

Taking the inverse Laplace transform over $\lambda$ only, we get for the Laplace transform of the hitting cost distribution $P(C|L)$
\begin{equation}
    \hat P(s|L)=\int_\Gamma \frac{d\lambda}{2\pi\mathrm{i}}\frac{1}{1-g(\lambda,s)}e^{\lambda L}\ ,\label{SMpole}
\end{equation}
where $\Gamma$ is a Bromwich contour. 

Assuming the large deviation form $P(C|L)\sim \exp(-L\Phi(C/L))$, how do we extract the rate function $\Phi(z)$ from the large-$L$ asymptotic behavior of $\hat P(s|L)$? The answer is provided by the Legendre-Fenchel theory as follows.
\begin{equation}
    \hat P(s|L)=\int_0^\infty dC e^{-s C}P(C|L)\approx
    \int_0^\infty dC e^{-s C}\exp(-L\Phi(C/L))=L\int_0^\infty dz \exp\left[-L (sz+\Phi(z))\right]\ ,
\end{equation}
where in the first step we have used the large deviation ansatz for $P(C|L)$, and then we changed variables $C/L=z$. For large $L$, we can evaluate the last integral via a saddle point method
\begin{equation}
    \hat P(s|L)\approx \exp\left[-L \min_z (sz+\Phi(z)) \right]\label{SMcomparison1}
\end{equation}
which should be compared with the predicted large deviation behavior
\begin{equation}
     \hat P(s|L)\approx \exp\left[L \lambda^\star (s)\right]\ ,\label{SMcomparison2}
\end{equation}
where the value $\lambda^\star(s)$ is that for which the integrand in \eqref{SMpole} has a pole, namely $g(\lambda^\star,s)=1$.

Comparing \eqref{SMcomparison1} and \eqref{SMcomparison2}, we get by Legendre-Fenchel inversion that
\begin{equation}
    \Phi(z)=\max_s\left[-sz+1+u(s)\right]\ ,
\end{equation}
namely Eq. (7) of the main text, after the identification $\lambda^\star(s)+1=-u(s)$.

We now investigate the behavior of the critical equation (8) of the main text
\begin{equation}
    bs+\left(bs e^s-1\right)u-u^2 e^s-bs e^{u\eta_c}=0\label{supp:eqforu}
\end{equation}
providing the location of the pole of Eq. \eqref{SMpole} after the identification $\lambda^\star(s)+1=-u(s)$. We first analyze it in the limit $s\to 0$. In this limit, the critical equation reduces to
\begin{equation}
    -u-u^2=0\Rightarrow u(s=0) = -1
\end{equation}
as the only nonzero root. Introducing a Taylor expansion ansatz $u(s)=-1+a_1 s-a_2 s^2$ for small $s$ into \eqref{supp:eqforu} and expanding for small $s$, we get
\begin{equation}
    s \left(a_1-b e^{-\eta_c }-1\right)+s^2 \left(-a_1^2-a_1 b e^{-\eta_c } \eta_c +a_1 b+2 a_1-a_2-b-\frac{1}{2}\right)+\mathcal{O}(s^3)=0\ .
\end{equation}
Equating the two coefficients to zero, we get
\begin{align}
    a_1 &=1+b e^{-\eta_c}\ ,\\
    a_2 &=\frac{1}{2}+b^2 e^{-\eta_c}(1-e^{-\eta_c})-b\eta_c e^{-\eta_c}(1+b e^{-\eta_c})\ .
\end{align}
Now, from Eq. (6) of the main text we get
\begin{align}
    \Phi(z) &=\max_s\left[-sz+1+u(s)\right]\\
    &=\max_s\left[-sz+a_1 s-a_2 s^2\right]
\end{align}
where the maximum is attained at $s^\star=(a_1-z)/2a_2$, where 
\begin{equation}
    \Phi(z)= -s^\star z+a_1 s^\star-a_2 s^{\star 2}=\frac{(z-a_1)^2}{4a_2}
\end{equation}
which corresponds to the quadratic minimum (central regime of eq. (9) in the main text).

\begin{figure}[t]
    \centering
    \textbf{\includegraphics[scale = 0.7]{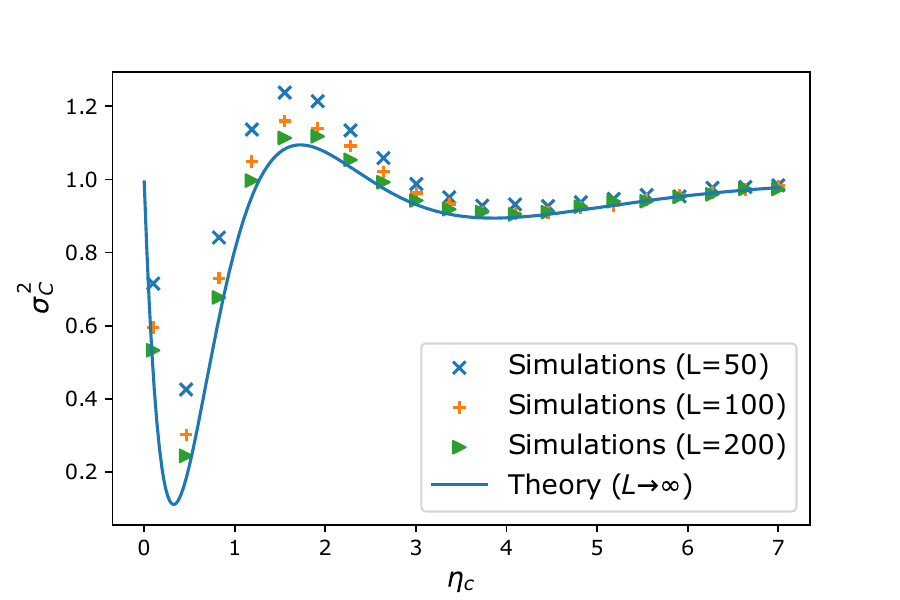}}
    \caption{Variance $\sigma_C^2$ of the hitting cost as a function of $\eta_c$ for $b=3.5$ and increasing values of $L$. Simulations are done averaging over $10^5$ trajectories that stop as soon as they cross the target spot $L$. In solid blue, the theoretical prediction in Eq. \eqref{varianceFPcost} of the main text.}
    \label{fig:sigmac_supmat}
\end{figure}

Coming back to $P(C|L)$, in the quadratic regime
\begin{align}
    P(C|L)\sim e^{-L\Phi(C/L)}\sim e^{-L \left(\frac{C}{L}-a_1\right)2}\frac{1}{4a_2}\sim e^{-\frac{(C-a_1 L)^2}{4a_2 L}}\ ,\label{supp:Gausscost}
\end{align}
which shows that the typical fluctuations of the hitting cost are Gaussian with mean and variance that can be read off from \eqref{supp:Gausscost} as
\begin{align}
    \langle C\rangle_L &= a_1 L = (1+b e^{-\eta_c})L\\
    \mathrm{Var}(C|L) &= \langle C^2\rangle_L-\langle C\rangle_L^2=2a_2 L = L\sigma_C^2\ ,
\end{align}
with $\sigma_C^2$ given in Eq.~\eqref{varianceFPcost} of the main text. The variance $\sigma_C^2$ is shown in Fig.~\ref{fig:sigmac_supmat} for different values of $L$. As a function of $\eta_c$, the variance $\sigma_C^2$ exhibits the rich behavior described in the main text.

We now analyze the critical equation \eqref{supp:eqforu} in the limit $s\to\infty$. We can first try to match the second and third term of the equation
\begin{equation}
    bs e^s u \simeq u^2 e^s\Rightarrow u\simeq bs\ ,\label{supp:matching1}
\end{equation}
but this can only hold true if the leading $s\to\infty$ asymptotics of the fourth term $bs e^{u\eta_c}$ (evaluated at $u\simeq bs$) turns out to be smaller than the $s\to\infty$ asymptotics of the second (or third) term $bs e^s u$ evaluated at $u\simeq bs$. Otherwise the initial assumption that the leading asymptotics be given by the second/third term of Eq. \eqref{supp:eqforu} would be internally inconsistent. 

So, we need to have $e^{bs\eta_c}<e^s$, which implies $b\eta_c<1$. In this sub-case,  setting $u=bs+y$ (where $y$ is a correction term), inserting this ansatz back into \eqref{supp:eqforu} we obtain
\begin{equation}
    -bs e^s y-y^2 e^s-bs e^{bs\eta_c}e^{y\eta_c}=0\Rightarrow y\simeq -e^{-s+bs\eta_c}=e^{-(1-b\eta_c)s}\label{eqfory}
\end{equation}
for large $s$. For $b\eta_c<1$, we have $y\ll 1$ for large $s$, which satisfies the equation \eqref{eqfory} self-consistently to leading order. Hence
\begin{equation}
    u(s)\simeq bs-e^{-(1-b\eta_c)s}
\end{equation}
when $b\eta_c<1$.

Now, from Eq. (6) of the main text we get
\begin{align}
    \Phi(z) &=\max_s\left[-sz+1+u(s)\right]\\
    &=\max_s\left[\underbrace{-sz+bs-e^{-(1-b\eta_c)s}}_{H(s)}\right]
\end{align}
where the maximum is attained at $s^\star$ that is solution of
\begin{equation}
    H'(s^\star)=b-z+(1-b\eta_c)e^{-(1-b\eta_c)s^\star}=0\Rightarrow s^\star = -\frac{1}{1-b\eta_c}\ln\left(\frac{z-b}{1-b\eta_c}\right)\ .
\end{equation}
Hence
\begin{equation}
    \Phi(z)=H(s^\star)=1+s^\star(b-z)-e^{-(1-b\eta_c)s^\star}=1+(z-b)\frac{1}{1-b\eta_c}\ln\left(\frac{z-b}{1-b\eta_c}\right)-\frac{z-b}{1-b\eta_c}
\end{equation}
as $z\to b^+$. Therefore, for $b<1/\eta_c$, we find that 
\begin{equation}
\psi_b(z)=1+(z-b)\frac{1}{1-b\eta_c}\ln\left(\frac{z-b}{1-b\eta_c}\right)-\frac{z-b}{1-b\eta_c}\,,
\end{equation}
and $\ell_1(b)=b$.

We now turn to the sub-case $b\eta_c>1$. In this case, the ``matching'' of terms in \eqref{supp:matching1} would be inconsistent. So, we need to match the second and fourth term in the critical equation \eqref{supp:eqforu}
\begin{equation}
    bse^s u\simeq b s e^{u\eta_c}\Rightarrow u\eta_c\simeq s +\ln u \Rightarrow u\simeq \frac{s}{\eta_c}+y\label{supp:uleading}
\end{equation}
where we can determine the correction term $y$ (to leading order in $s$) by substituting \eqref{supp:uleading} into \eqref{supp:eqforu}
\begin{align}
    & bs+(bs e^s-1)(s/\eta_c+y)-e^s (s^2/\eta_c^2+2y s/\eta_c+y^2)-bs  e^s e^{y\eta_c}=0 \Rightarrow\\
    & bs+\frac{bs^2}{\eta_c}e^s+bs e^s y
-\frac{s}{\eta_c}-y-\frac{s^2}{\eta_c^2}e^s-2y\frac{s}{\eta_c}e^s-y^2 e^s-bs e^s e^{y\eta_c}=0\ .\end{align}

Lumping together the leading terms ($\sim s^2 e^2$) on the left hand side, we need to have
\begin{equation}
    \left(\frac{b}{\eta_c}-\frac{1}{\eta_c^2}\right)s^2 e^s\simeq bs e^s e^{y\eta_c}
\end{equation}
from which
\begin{equation}
    y\simeq \frac{1}{\eta_c}\ln\left(\frac{b\eta_c-1}{b\eta_c^2}s\right)\ .
\end{equation}
Substituting $u(s)\simeq s/\eta_c+\frac{1}{\eta_c}\ln\left(\frac{b\eta_c-1}{b\eta_c^2}s\right)$ into Eq. (6) of the main text
\begin{align}
    \Phi(z) &=\max_s\left[-sz+1+u(s)\right]\\
    &=\max_s\left[\underbrace{-sz+1+\frac{s}{\eta_c}+\frac{1}{\eta_c}\ln s +\mathcal{O}(1)}_{K(s)}\right]
\end{align}
where the maximum is attained at $s^\star$ that is solution of
\begin{equation}
    K'(s^\star)=-z+1/\eta_c+\frac{1}{\eta_c s^\star}=0\Rightarrow s^\star = \frac{1}{\eta_c \left(z-\frac{1}{\eta_c}\right)}\ .
\end{equation}
Hence
\begin{equation}
    \Phi(z)=K(s^\star)=\frac{1}{\eta_c}\ln\left[\frac{1}{\eta_c \left(z-\frac{1}{\eta_c}\right)}+(1/\eta_c-z)\frac{1}{\eta_c \left(z-\frac{1}{\eta_c}\right)}+\mathcal{O}(1)\right]=-\frac{1}{\eta_c}\ln\left(z-\frac{1}{\eta_c}\right)+\mathcal{O}(1)\ ,
\end{equation}
which diverges logarithmically as $z\to (1/\eta_c)^+$. Thus, for $b>1/\eta_c$, we find 
\begin{equation}
\psi_b(z)=-\frac{1}{\eta_c}\ln\left(z-\frac{1}{\eta_c}\right)+\mathcal{O}(1)\,,
\end{equation}
and $\ell_1(b)=1/\eta_c$.

We can now analyze the behavior exactly on the transition line, namely when $b=1/\eta_c$ exactly. In this case, the critical equation \eqref{supp:eqforu} reads
\begin{equation}
    bs+\left(bs e^s-1\right)u-u^2 e^s-bs e^{u/b}=0\ .\label{supp:eqforucriticalline}
\end{equation}
Matching again the second and third term of the equation
\begin{equation}
    bs e^s u \simeq u^2 e^s\Rightarrow u\simeq bs\ ,\label{supp:matching1criticalline}
\end{equation}
and assuming therefore $u\simeq bs + \tilde{y}$, the correction term $\tilde{y}$ satisfies
\begin{align}
  & bs+\left(bs e^s-1\right)(bs+\tilde{y})-(bs+\tilde{y})^2 e^s-bs e^{s+\tilde{y}/b}=0  \Rightarrow\\
  & -bs e^s\tilde{y}-\tilde{y}-e^s\tilde{y}^2-bs e^s e^{\tilde{y}/b}=0
\end{align}
and assuming the behavior $\tilde{y}\simeq \delta_0+\delta_1/s$ of the correction term, we get in the limit $s\to\infty$ the self-consistency conditions for the constants $\delta_0$ and $\delta_1$
\begin{align}
&-bs e^s(\delta_0+\delta_1/s)-(\delta_0+\delta_1/s)-e^s(\delta_0+\delta_1/s)^2-bs e^s e^{(\delta_0+\delta_1/s)/b}=0\Rightarrow \\
& bs e^s \left(-\delta_0-e^{\delta_0/b}\right)+e^s\left(-b\delta_1-\delta_0^2-\delta_1 e^{\delta_0/b}\right)=0\,,
\label{eq:large_s_crit}
\end{align}
where in the last step we expanded $\exp\left(\frac{\delta_1}{sb}\right)\approx 1+\frac{\delta_1}{sb}$ and neglected sub-leading terms for large $s$. In the limit $s\to\infty$, the terms in parentheses in Eq.~\eqref{eq:large_s_crit} must vanish, leading to the trascendental equations
\begin{equation}
    \delta_0=-e^{\delta_0/b}\,,
    \label{eq:delta_0_condition}
\end{equation}
and
\begin{equation}
    \delta_1=-\frac{\delta_0^2}{b-\delta_0}\,.
    \label{eq:delta_1_condition}
\end{equation}
Since $b>0$, one finds $\delta_0<0$ and $\delta_1<0$. Therefore, we find that the solution $u(s)$ of Eq.~\eqref{supp:eqforu}, can be written as, for $s\to\infty$
\begin{equation}
    u(s)\approx bs+\delta_0+\frac{\delta_1}{s}\,.
\end{equation}
Inserting this result into the equation for the rate function (Eq. (6) of the main text), we find that for small $z$
\begin{equation}
    \Phi(z)\approx \max_s\left(-sz+1+bs+\delta_0+\frac{\delta_1}{s}\right)\,.
\end{equation}
Performing the maximization over $s$, we obtain that, for $z>b$,
\begin{equation}
    \Phi(z)\approx -2\sqrt{-\delta_1(z-b)}+1+\delta_0\,.
\end{equation}

Thus, for the critical case $b=1/\eta_c$, we find
\begin{equation}
\psi_b(z)=-2\sqrt{-\delta_1(z-b)}+1+\delta_0\,,
\end{equation}
and $\ell_1(b)=b$, as reported in the main text.

What remains to be done is to analyze the behavior as $s\to -\infty$ of the critical equation \eqref{supp:eqforu}. As $s\to -\infty$, the dominant terms to be matched are
\begin{equation}
    -u-u^2 e^s\simeq 0\Rightarrow u(s)\simeq -e^{-s}
\end{equation}
which works self-consistently in \eqref{supp:eqforu}, as the first and second terms ($bs$ and $bs e^s u$) would then cancel out exactly, while the last term ($-bs e^{u\eta_c}$) would go to zero as $s\to -\infty$.

Substituting $u(s)\simeq -e^{-s}$ into Eq. (6) of the main text
\begin{align}
    \Phi(z) &=\max_s\left[\underbrace{-sz+1-e^{-s}}_{F(s)}\right]
\end{align}
where the maximum is attained at $s^\star$ that is solution of
\begin{equation}
    F'(s^\star)=0\Rightarrow -z+e^{-s^\star}=0\Rightarrow s^\star = -\ln z
\end{equation}
from which
\begin{equation}
    \Phi(z)=F(s^\star)=z\ln z+1-z
\end{equation}
as $z\to +\infty$ (corresponding to the asymptotics $s\to -\infty$).

\subsection{Ensemble (ii): Configuration of minimal cost and the freezing transition}

In this Section, we identify the trajectories that minimize the total cost $C$ for Ensemble (ii) in the limit $L\to\infty$. This minimization will allow us to understand the freezing transition of the lower edge $\ell_1(b)$ of the rate function $\Phi(z)$ around Eq. (11) of the main text. Indeed, the lower edge $\ell_1(b)$ is related to the minimal cost $C_{\rm min}$ by the relation $C_{\rm min}=\ell_1(b)L$.

Assuming that the trajectory of minimal cost is composed of $n$ jumps of the same length $\eta=L/n$, we have that the total cost is
\begin{equation}
    C=n h\left(\frac{L}{n}\right)\,,
\end{equation}
where $h(\eta)=1+b(\eta-\eta_c)\theta(\eta-\eta_c)$ is the single-step cost function. The total cost can be rewritten as
\begin{equation}
    C=n +b(L-\eta_c n)\theta(L-\eta_cn)\,.
\end{equation}

Minimizing the cost with respect to $n$, we find that for $b<1/\eta_c$ the minimal cost is 
\begin{equation}
    C_{\rm min}=bL\,,
\end{equation}
corresponding to a single jump ($n=1$) of length $\eta=L$. In other words, if space-like jump are sufficiently cheap (i.e., if $b$ is small), the minimal cost is attained by a single long jump. On the other hand, for $b>1/\eta_c$ we find
\begin{equation}
    C_{\rm min}=L/\eta_c\,,
\end{equation}
corresponding to $n=L/\eta_c$ time-like steps of maximal length $\eta=\eta_c$. This explains the freezing transition of the lower edge $\ell_1(b)$.

\section{Generalized cost function}

As explained in the main text, our results can be applied to a range of physical systems that
typically undergo a depinning transition when driven by an external force. In these systems, an extended
object such as an elastic string in a disordered medium is dragged by an external force $\eta$, it remains immobile for
$\eta<\eta_c$, and starts sliding with a constant velocity $v\propto (\eta-\eta_c)^{\beta}$ with
$\beta>0$ when the applied force exceeds the threshold value $\eta_c$. The exponent $\beta$ depends
on the system. When a random force (say, exponentially distributed)
is now applied to such an extended object, it is natural to investigate what the sample averaged velocity-force
characteristic looks like. Fluctuations around this average response is also important to understand.
Out taxi random walk model is precisely suitable to address this question. In this model we
had chosen the cost function to be $h(\eta)= 1+ b (\eta-\eta_c)\theta(\eta-\eta_c)$. To apply
our results to the randomly forced elastic string mentioned above, we just need to generalize our computations
to the case $h(\eta)= b(\eta-\eta_c)^{\beta}\theta(\eta-\eta_c)$ where $\beta>0$ is arbitrary.
In this section, we present an exact computation for the class of cost functions
\begin{equation}
h(\eta)=1+b(\eta-\eta_c)^{\beta}\theta(\eta-\eta_c)\,,
\label{h_beta}
\end{equation}
for arbitrary $\beta>0$, thus generalizing our previous result for
$\beta=1$. Note that we have included the constant shift $1$ in Eq. (\ref{h_beta}) in order to compare
to the taxi random walk model for $\beta=1$. In the physics examples mentioned above, one does not need
this extra $1$ factor. However, the contribution of this extra factor $1$ in $h(\eta)$ is just $n$ in the total cost
and it does not affect the variance.
We will focus here on Ensemble (i), i.e., fixing $X$ and $n$ and investigate the first two moments of 
the cost.

\begin{figure}[t]
    \centering
    \textbf{\includegraphics[scale = 0.5]{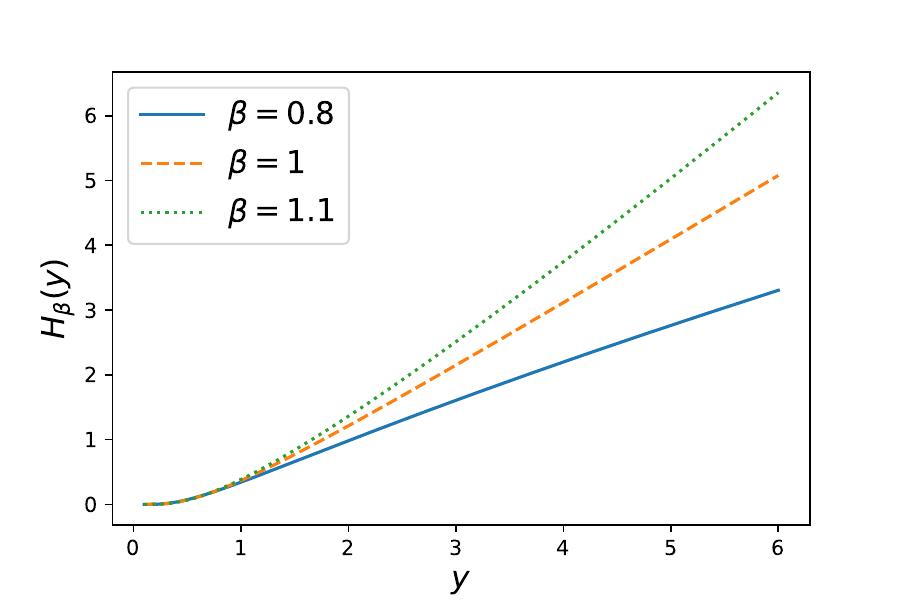}}
    \textbf{\includegraphics[scale = 0.5]{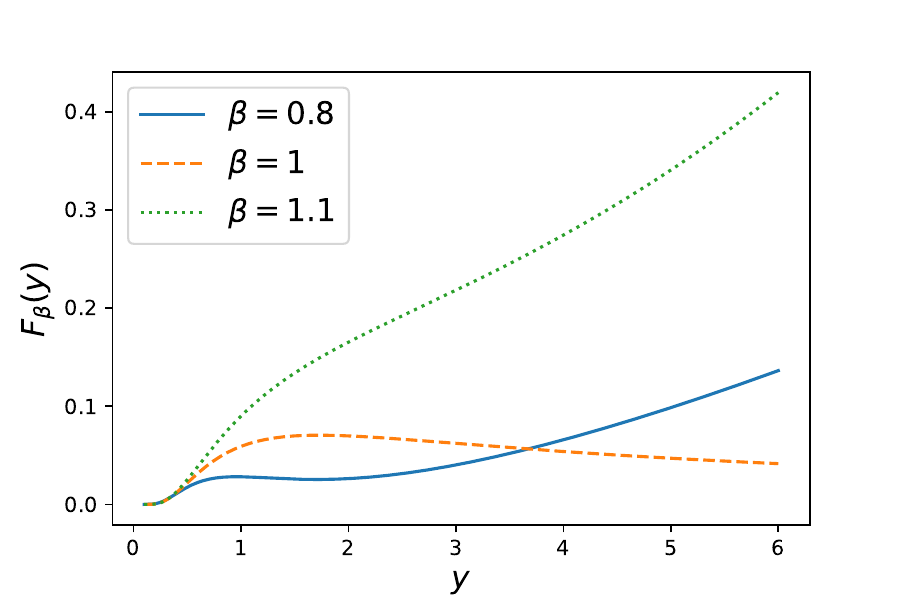}}
    \caption{Scaling functions $H_{\beta}(y)$ (left panel) and $F_{\beta}(y)$ (right panel), respectively given in Eqs.~\eqref{eq:sm_hbeta} and \eqref{eq:sm_fbeta}, as a function of $y$ for different values of $\beta$.}
    \label{fig:HF}
\end{figure}

Analogously to the case $\beta=1$, the $k$-th moment of the cost can be computed by 
first evaluating (see Eq.~\eqref{SM:derivatives})
\begin{equation}
\int_0^{\infty} \int_0^{\infty} dC~ C^k P(C,X|n) e^{-\lambda X} dX
= (-1)^k \frac{d^k}{d s^k} [g(\lambda,s)]^n \Big|_{s=0}\ ,
\label{eq:Ck}
\end{equation}
where 
\begin{equation}
    g(\lambda, s)=\int_0^\infty d\eta~p(\eta)e^{-\lambda \eta-s h(\eta)}\,.
\end{equation}
Using the expression in Eq.~\eqref{h_beta}, we find
\begin{equation}
g(\lambda,s)=e^{-s}\left[\frac{1}{\lambda+1}\left(1-e^{-(\lambda+1)\eta_c}\right)+e^{-(\lambda+1) \eta_c}\int_{0}^{\infty}d\eta~e^{-(\lambda+1)\eta-sb(\eta-\eta_c)^{\beta}}\right]\,.
\end{equation}
From Eq.~\eqref{eq:Ck} with $k=1$, we obtain, after a few steps of algebra,
\begin{equation}
\int_0^{\infty} \int_0^{\infty} dC~ C P(C,X|n) e^{-\lambda X} dX
= n\frac{1}{(\lambda+1)^{n-1}}\left[\frac{1}{\lambda+1}+\frac{b\Gamma(\beta+1)}{(\lambda+1)^{\beta+1}}e^{-\lambda\eta_c}\right]\ .
\end{equation}
By inverting the Laplace transform with respect to $X$ using Eq.~\eqref{eq:Laplace_inversions}, we get
\begin{equation}
 \int_0^{\infty} dC~ C P(C,X|n) 
= ne^{-X}\left[\frac{X^{n+1}}{\Gamma(n)}+\frac{b\Gamma(\beta+1)}{\Gamma(\beta+n)}(X-\eta_c)^{\beta+n-1}\theta(X-\eta_c)\right]\,.
\end{equation}
Finally, dividing by $P(X|n)$, given in Eq.~\eqref{eq:PXNsm}, we find
\begin{equation}
\langle C\rangle_{X,n}=n+bn\frac{\Gamma(\beta+1)\Gamma(n)}{\Gamma(\beta+n)}\left(X-\eta_c\right)^{\beta+n-1}\theta(X-\eta_c)X^{-(n-1)}\ .
\end{equation}
Note that by setting $\beta=1$ we recover the result in Eq.~\eqref{conditionalexpectation}. In the limit $X\to\infty$, $n\to \infty$ with $X/n$ fixed the average cost can be written as
\begin{equation}
\langle C\rangle_{X,n}=n+b\eta_c^{\beta}n H_{\beta}\left(\frac{X}{\eta_c n}\right)\,,
\label{mean_cost_beta}
\end{equation}
where
\begin{equation}
H_{\beta}(y)=\Gamma(\beta+1)y^{\beta}e^{-1/y}\,.
\label{eq:sm_hbeta}
\end{equation}
This scaling function $H_{\beta}(y)$ is shown in Fig.~\eqref{fig:HF}. In physical systems, the shift factor $n$ in Eq. (\ref{mean_cost_beta}) will not be there and for large $n$, the scaling function
$H_{\beta}(y)$ then describes precisely the average velocity-force characteristic (per sample) when
the extended object is driven by an external random force $\eta$ drawn from an exponential
distribution $p(\eta)= e^{-\eta}$.

We next focus on the cost variance. Following the same procedure as 
above and using the Laplace inversion formula in Eq.~\eqref{eq:Laplace_inversions}, we find
\begin{align}
\operatorname{Var}(C|X,n)&=n(n-1)b^2\frac{\Gamma^2(\beta+1)\Gamma(n)}{\Gamma(n+2\beta)}\frac{(X-2\eta_c)^{n+2\beta-1}}{X^{n-1}}\theta(X-2\eta_c)+b^2 n \frac{\Gamma(2\beta+1)\Gamma(n)}{\Gamma(n+2\beta)}\frac{(X-\eta_c)^{n+2\beta-1}}{X^{n-1}}\theta(X-\eta_c)\\
\nonumber &-b^2\frac{\Gamma^2(\beta+1)\Gamma^2(n)}{\Gamma^2(\beta+n)}n^2\frac{(X-\eta_c)^{2(\beta+n-1)}}{X^{2(n-1)}}\theta(X-\eta_c)\,.
\end{align}
Taking the limit $n\to\infty$, $X\to \infty$ with $X/n$ fixed we find
\begin{equation}
\operatorname{Var}(C|X,n)\approx nb^2\eta_c^{2\beta}F_{\beta}\left(\frac{X}{\eta_c n}\right)\,,
\end{equation}
where 
\begin{equation}
F_{\beta}(y)= y^{2\beta}\left[\Gamma(2\beta+1)e^{-1/y}-(\beta^2+1)\Gamma^2(\beta+1)e^{-2/y}-2\beta\frac{1}{y}e^{-2/y}\Gamma^2(\beta+1)-\frac{1}{y^2}e^{-2/y}\Gamma^2(\beta+1)\right]\,.
\end{equation}
Note that for $\beta=1$ we recover the result in Eq.~\eqref{Fysm}. The scaling function $F_{\beta}(y)$ is plotted in Fig.~\eqref{fig:HF} and
has asymptotic behaviors
\begin{equation}
F_{\beta}(y)\approx\begin{cases}
\Gamma(2\beta+1)y^{2\beta}e^{-1/y}\,,\quad & \text{ for } y\to 0\,,\\
\\
\left[\Gamma(2\beta+1)-(1+\beta^2)\Gamma(\beta+1)^2\right]y^{2\beta}\,,\quad &\text{ for } y\to \infty\,.
\end{cases}
\label{eq:sm_fbeta}
\end{equation}
In the large-$y$ limit the cost of the cost variance grows as $y^{2\beta}$ for $\beta\neq 1$. However, 
only for $\beta=1$,
the prefactor of the leading term $y^{2\beta}$ for large $y$ vanishes exactly and the first nonzero term turns
out to be $1/(3y)$, 
leading to vanishing fluctuations for large $y$. As explained in the main text, this is a consequence of the 
linearity of $h(\eta)$ for large $\eta$ in the case $\beta=1$.

\vspace{0.5cm}

\end{document}